\documentclass[12pt,twocolumn]{elsart}

\usepackage{graphics}
\usepackage{amssymb}

\newcommand{\vek}[1]{\mbox{\bf #1}}
\def\WIMPZILLA{{\sc wimpzilla}}
\def\WIMPZILLAS{{\sc wimpzillas}}
\def\WIMP{{\sc wimp}}
\def\WIMPS{{\sc wimps}}

\begin{document}

\begin{frontmatter}

\title{Ultra-high energy cosmic rays from\\ 
annihilation of superheavy dark matter}

\author[label1,label2]{Pasquale Blasi\thanksref{pasquale}},
\author[label3]{Rainer Dick\thanksref{rainer}}, and
\author[label2,label4]{Edward W. Kolb\thanksref{rocky}}

\address[label1]{Osservatorio Astrofisico di Arcetri, 
Largo E. Fermi 5, 50125 Firenze, Italy}
\address[label2]{NASA/Fermilab Astrophysics Center,
        Fermi National Accelerator Laboratory,\\ Batavia, Illinois 60510-0500}
\address[label3]{Department of Physics and Engineering Physics,
        University of Saskatchewan,\\ 116 Science Place, Saskatoon, 
        SK S7N 5E2, Canada}
\address[label4]{Department of Astronomy and Astrophysics,
        Enrico Fermi Institute,\\ The University of Chicago, Chicago, Illinois 
        60637-1433}
\thanks[pasquale]{blasi@arcetri.astro.it} %New address!
\thanks[rainer]{rainer@sask.usask.ca}
\thanks[rocky]{rocky@rigoletto.fnal.gov,\\
 \hspace*{2.5mm} Rocky.Kolb@cern.ch}

\begin{abstract}
We consider the possibility that ultra-high energy cosmic rays originate from
the annihilation of relic superheavy dark matter.  We find that a cross section
of $\langle\sigma_Av\rangle\sim 10^{-26}\textrm{cm}^2
(M_X/10^{12}\,\textrm{GeV})^{3/2}$ is required to account for the observed rate
of super-GZK events if the superheavy dark matter follows a
Navarro--Frenk--White density profile.  This would require extremely large-$l$
contributions to the annihilation cross section.  We also calculate the
possible signature from annihilation in sub-galactic clumps of dark matter and
find that the signal from sub-clumps dominates and may explain the observed
flux with a much smaller cross section than if the superheavy dark matter is
smoothly distributed.  
Finally, we discuss the expected anisotropy in the arrival directions
of the cosmic rays, which is a characteristic signature of this scenario.
\end{abstract}


\end{frontmatter}

%%%%%%%%%%%%%%%%%%%%%%%%%%%%%%%%%%%%%%%%%%%%%%%%%%%%%%%
%%%%%%%%%%%%%%%%%%%%%%%%%%%%%%%%%%%%%%%%%%%%%%%%%%%%%%%
\section{Introduction}
%%%%%%%%%%%%%%%%%%%%%%%%%%%%%%%%%%%%%%%%%%%%%%%%%%%%%%%
%%%%%%%%%%%%%%%%%%%%%%%%%%%%%%%%%%%%%%%%%%%%%%%%%%%%%%%

The nature of the dark matter and the origin of the ultra-high energy (UHE)
cosmic rays are two of the most pressing issues in contemporary particle
astrophysics.  In this paper we explore a possible connection between these
issues: dark matter is a very massive relic particle species (\WIMPZILLA), and
\WIMPZILLA\ annihilation products are the ultra-high energy cosmic rays.

The dark matter puzzle results from the observation that visible matter can
account only for a small fraction of the matter bound in large-scale
structures.  The evidence for dark matter is supported by mass estimates from
gravitational lensing \cite{lens}, by the peculiar velocities of large scale
structures \cite{flow}, by measurements of CMB anisotropy \cite{CMB} and by
measurements of the recession velocity of high-redshift supernovae
\cite{supernovae}.  Constraints from big-bang nucleosynthesis imply that the
bulk of the 
dark matter cannot be baryonic, and most of the matter density of the universe
must arise from particles not accounted for by the Standard Model of particle
physics \cite{BBN}.

The existence of UHE cosmic rays of energies above the Greisen-Zatsepin-Kuzmin
cutoff \cite{GZK}, $E_{\mathrm{GZK}}\simeq 5\times 10^{19}\,$eV, is a major
puzzle because the cosmic microwave background constitutes an efficient
obstacle for protons or nuclei of ultra-high energies to travel farther than a
few dozen Mpc \cite{GZK,new}.  This suggests that the observed extremely
energetic cosmic rays with $E>E_{\mathrm{GZK}}$ should originate in our cosmic
neighborhood. Furthermore, the approximately isotropic distribution of arrival
directions makes it difficult to imagine that nearby astrophysical sources are
the accelerators of the observed UHE cosmic rays (but for the opposite point of
view, see \cite{gftp}).

An interesting possibility is that UHE cosmic rays arise from the
decay of some superheavy particle.  This possibility has been proposed 
by Berezinsky, Kachelrie\ss\ and Vilenkin
and by Kuzmin and Rubakov \cite{BKV,KR},
see also \cite{JELLIS} for a discussion of this in the framework
of string/M-theory, and \cite{bbv} for a discussion in the framework
of topological defects. The superheavy particles
must have masses $M_X\ge 10^{12}\,$GeV.  Although this proposal circumvents
 the astronomical problems, there are two issues to address: Some
cosmological mechanism must be found for producing particles of such large mass
in the necessary abundance, and the lifetime of this very massive state must be
in excess of $10^{20}\,$yr, if we want these particles to be both dark matter
candidates and sources of UHE cosmic rays.

The simple assumption that dark matter is a thermal relic limits the maximum
mass of the dark matter particle.  The largest annihilation cross section in
the early universe is expected to be roughly $M_X^{-2}$.  This implies that
very massive \WIMPS\ have such a small annihilation cross section that their
present abundance would be too large if the \WIMPS\ are thermal relics.  Thus,
one predicts a maximum mass for a thermal \WIMP, which turns out to be a few
hundred TeV.  While a thermal origin for \WIMPS\ is the most common assumption,
it is not the simplest possibility.  It has been recently pointed out that dark
particles might have never experienced local chemical equilibrium during the
evolution of the universe, and that their mass may be in the range $10^{12}$ to
$10^{19}$ GeV, much larger than the mass of thermal \WIMPS\
\cite{ckr1,ckr2,ckr3,ckr4}. Since these \WIMPS\ would be much more massive than
thermal \WIMPS, such superheavy dark particles have been called \WIMPZILLAS\
\cite{ckr4}.

Since \WIMPZILLAS\ are extremely massive, the challenge lies in creating very
few of them. Several \WIMPZILLA\ scenarios have been developed involving
production during different stages of the evolution of the universe. 

\WIMPZILLAS\ may be created during bubble collisions if inflation is completed
through a first-order phase transition \cite{barrow,mr}; at the preheating
stage after the end of inflation with masses easily up to the Grand Unified
scale of $10^{15}$GeV \cite{Kolb:1998he} or even up to the Planck scale
\cite{danthesis,CKRT,Giudice:1999fb}; or during the reheating stage after 
inflation \cite{ckr3} with masses which may be as large as $2\times 10^3$ times
the reheat temperature.

\WIMPZILLAS\ may also be generated in the transition between an inflationary
and a matter-dominated (or radiation-dominated) universe due to the
``nonadiabatic'' expansion of the background spacetime acting on the vacuum
quantum fluctuations. This mechanism was studied in details in Refs.\
\cite{ckr1,kuzmin,cckr}. The distinguishing feature of this mechanism is the 
capability of generating particles with mass of the order of the inflaton mass
(usually much larger than the reheat temperature) even when the particles only
interact extremely weakly (or not at all) with other particles, and do not
couple to the inflaton.

The long lifetime required in the decay scenario for UHE cosmic rays is
problematic if the UHE cosmic rays originate in single-particle decays.  The
lifetime problem can be illustrated in the decays of string or Kaluza--Klein
dilatons.  These particles 
can be decoupled from fermions \cite{rdgrg}, 
and therefore decay in leading order through their dimension-five 
couplings to gauge fields
\begin{equation}
\label{couple}
\mathcal{L}_{I\Phi}=-\frac{\Phi}{4f}\ F^{\mu\nu}{}_aF_{\mu\nu}{}^a.
\end{equation}
Here, $f$ is a mass scale characterizing the strength of the coupling, and in
Kaluza--Klein or string theory, it is of order of the reduced Planck mass
$m_{\mathrm{Pl}}=(8\pi G_N)^{-1/2}\simeq 2.4\times 10^{18}\,$GeV.  In heterotic
string theory, {\it e.g.,} $f=m_{\mathrm{Pl}}/\sqrt{2}\simeq 1.7\times
10^{18}\,$GeV \cite{rd2}.  If the number of the vector fields is $d_G$, we find
the lifetime estimate from the dilaton--vector coupling of Eq.\ (\ref{couple})
\begin{eqnarray*}
\tau_{\Phi}& =& \frac{32\pi f^2}{d_G m_{\phi}^3} \nonumber \\
 & = & \frac{1.9\times 10^{-22}\,\mbox{s}}{d_G} 
\left(\frac{10^{12}\,\mbox{GeV}}{m_{\phi}}\right)^3 
\nonumber \\ & & \times \left(\frac{f}{1.7\times 10^{18}\,\mbox{GeV}}\right)^2.
\end{eqnarray*}
If there are direct decay channels through first order couplings,
superheavy particles decay extremely fast, 
even if the coupling is only of gravitational strength 
and dimensionally suppressed.
Superheavy relic particles with a sufficiently long lifetime therefore require
sub-gravitational couplings or exponential suppression of the decay mechanism
due to wormhole effects \cite{BKV}, instantons \cite{KR}, or magic from the
brane world \cite{PAULF}.

Motivated by the attractiveness of the decay scenario, we investigate the
possibility that UHE cosmic rays may result from annihilation of relic
superheavy dark matter.

Annihilation of dark matter in the halo has been a subject of much interest,
with particular emphasis on possible neutralino signatures in the cosmic ray
flux. A reasonable annihilation scenario for UHE cosmic rays is the production
of two jets, each of energy $M_X$, which then fragment into a
many-particle final state including leading particles of energies 
comparable to a significant fraction of $M_X$. 

Assuming that the relic superheavy dark matter follows a Navarro--Frenk--White
(NFW) density profile \cite{NFW}, in Sec.\ \ref{smooth} we calculate the
expected spectrum and the annihilation cross section required to account for
the observed super-GZK events. We find that the necessary cross section
exceeds the unitarity bound unless the annihilation involves very large
angular momentum contributions or non-standard mechanisms, like, e.g.,
in monopolonium decay.

In Sec.\ \ref{clumped} we consider the contribution from annihilation in
sub-galactic clumps of dark matter.  We find that 
annihilation in isothermal sub-galactic clumps may explain the UHE cosmic rays
without violating the unitarity bound.

In Sec.\ \ref{anisotropy} we discuss the expected anisotropy in arrival
direction of UHE cosmic rays if the primary source is superheavy relic particle
annihilation in sub-galactic clumps. 

Sec.\ \ref{conclusion} contains our conclusions.

%%%%%%%%%%%%%%%%%%%%%%%%%%%%%%%%%%%%%%%%%%%%%%%%%%%%%%%
%%%%%%%%%%%%%%%%%%%%%%%%%%%%%%%%%%%%%%%%%%%%%%%%%%%%%%%
\section{Annihilation in the smooth component}
\label{smooth}
%%%%%%%%%%%%%%%%%%%%%%%%%%%%%%%%%%%%%%%%%%%%%%%%%%%%%%%
%%%%%%%%%%%%%%%%%%%%%%%%%%%%%%%%%%%%%%%%%%%%%%%%%%%%%%%

In calculating the UHE cosmic ray flux from a smooth\footnote{We refer to a
superheavy dark matter distribution as {\it smooth} if it can be described by a
particle density $n_X(\vek{d})$ which decreases uniformly with distance from
the galactic center.} superheavy dark matter distribution in the galactic halo,
we assume a superheavy $X$-particle halo density spherically symmetric about
the galactic center, $n_X(\vek{d})=n_X(d)$, where $d$ is the distance from the
galactic center.  We will assume $n_X(d)$ is given by an NFW profile \cite{NFW}
\begin{equation}
\label{nfwp}
n_X(d)=\frac{N_0}{d(d+d_s)^2} .
\end{equation}

Navarro's estimate for the fiducial radius $d_s$ for the Milky Way is of order
$25\,$kpc \cite{julio}.  Dehnen and Binney have examined a flattened NFW
profile as a special case of a whole class of halo models and give a value
$d_s=21.8\,$kpc \cite{DB}.  We will use $d_s = 3d_\odot = 24\,$kpc in our
numerical estimates, where $d_\odot\simeq8\,$kpc is the distance of the solar
system from the galactic center.

The dimensionless parameter $N_0$ may be found by requiring that the total mass
of the Galaxy is $2\times 10^{12} M_{\odot}$, which gives
\[
N_0 = \frac{8\times10^{55}}{M_{12}} = \frac{2.8\times10^{-9}}{M_{12}}
              \frac{\textrm{kpc}^3}{\textrm{cm}^3} ,
\] 
where $M_{12}=M_X/10^{12}\,\textrm{GeV}$.

For simplicity, we will assume that \WIMPZILLA\ annihilation produces two
leading jets, each of energy $M_X$, while decay of a \WIMPZILLA\ produces two
jets, each of energy $M_X/2$.  The energy spectrum of observed UHE cosmic ray
events from annihilation is
\begin{eqnarray}
\label{smoothannihilation}
\mathcal{F} &  = & 2 \ \frac{d\mathcal{N}(E,E_\textrm{\small jet}=M_X)}{dE} \  
\langle\sigma_Av\rangle \nonumber \\ & & \times
\int d^3\!\vek{d}\ \frac{n_X^2(d)}{4\pi \left|\vek{d}-\vek{d}_\odot\right|^2} .
\end{eqnarray}
Here, $d\mathcal{N}(E,E_\textrm{\small jet})/dE$ is the fragmentation spectrum
resulting from a jet of energy $E_\textrm{\small jet}$.  For comparison, the energy
spectrum of observed UHE cosmic ray events from \WIMPZILLA\ decay is
\begin{eqnarray*}
\mathcal{F} &  = &  2 \ \frac{d\mathcal{N}(E,E_\textrm{\small jet}=M_X/2)}{dE} \ 
\Gamma_X \nonumber \\ & & \times
\int d^3\!\vek{d}\ \frac{n_X(d)}{4\pi \left|\vek{d}-\vek{d}_\odot\right|^2} ,
\end{eqnarray*}
where $\Gamma_X$ is the decay width of the \WIMPZILLA.

Extrapolations and approximations to the
fragmentation function $d\mathcal{N}(E,E_\textrm{\small jet})/dE$
have been
reviewed in \cite{BS}. Birkel and Sarkar applied the {\sc Herwig}
Monte Carlo program to calculate the spectrum from \WIMPZILLA\ decay
\cite{BSark}, and Fodor and Katz employed a numerical integration
of the DGLAP evolution equations \cite{FK}.
The resulting spectrum in the interesting range
between $10^{19}\,$GeV and $10^{20}\,$GeV is similar to the spectrum
from the modified leading
logarithmic approximation (MLLA) \cite{MLLA}, which was
employed by Berezinsky et al.\ in their proposal of UHE cosmic rays
from decay of superheavy dark matter decay \cite{BKV}.

\begin{figure}[htb]
\begin{center}
\resizebox{0.5\textwidth}{!}{ \includegraphics{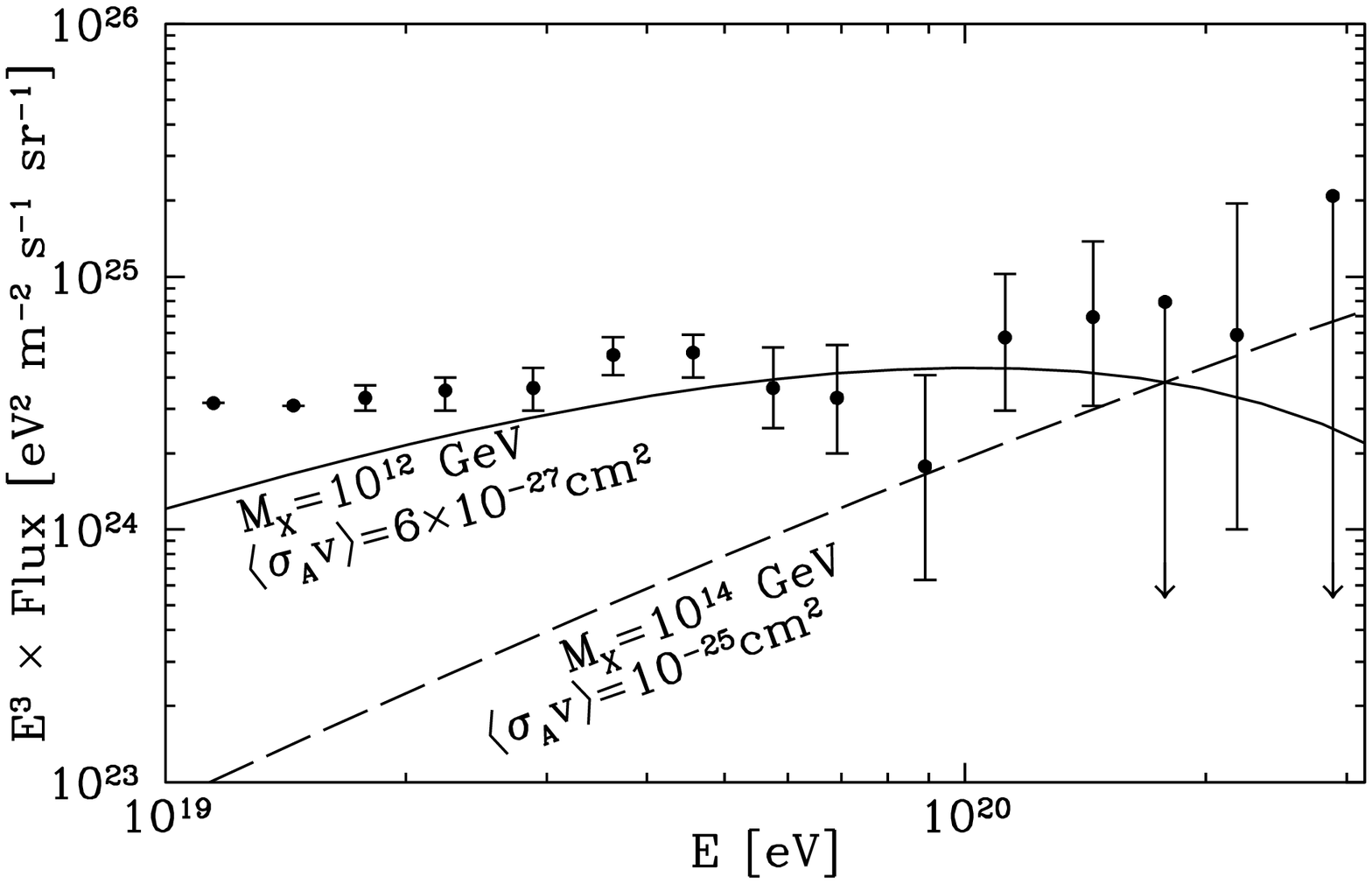} }
\resizebox{0.5\textwidth}{!}{ \includegraphics{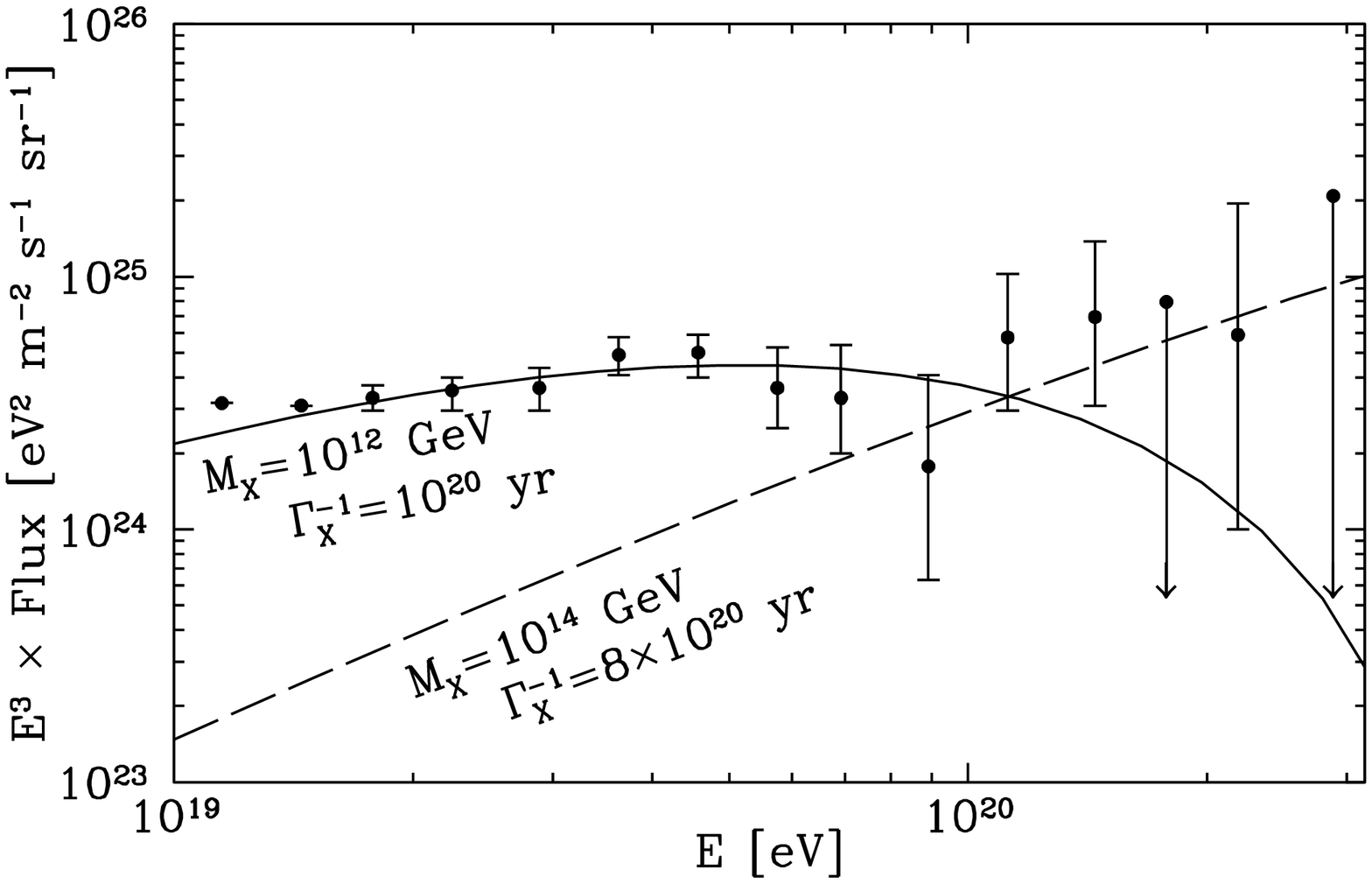} }
\end{center}
\caption{UHE cosmic ray spectra from superheavy particle annihilation (upper
panel) or decay (lower panel).  For both figures the solid lines are for
$M_X=10^{12}\,$GeV and the dashed lines are for $M_X=10^{14}\,$GeV.  For
annihilation, the solid curve is for $\langle\sigma_Av\rangle = 
6\times10^{-27}\,\textrm{cm}^2$ and the dashed curve corresponds to  
$\langle\sigma_Av\rangle = 10^{-25}\,\textrm{cm}^2$.  In the decay case, the
solid curve is for $\Gamma_X^{-1}=10^{20}\,$yr and the dashed curve is for
 $\Gamma_X^{-1} = 8\times10^{20}\,$yr.}
\label{spectrum}     
\end{figure}

We also use the MLLA limiting spectrum in the
results of Fig.\ \ref{spectrum}.
The salient features
of the results can be understood by making the simple approximation that most
of the content of the jet is pions, with a spectrum in terms of the usual
variable $x \equiv E/E_\textrm{jet}$ (of course, $0 \leq x \leq 1$),
\begin{eqnarray*}
 \frac{d\mathcal{N}(x)}{dx} & = & 
\frac{15}{16}x^{-3/2}(1-x)^2 \nonumber \\
& \sim & \frac{15}{16}x^{-3/2}\ \ (x\ll 1) .
\end{eqnarray*}

Using this fragmentation approximation, the scaling of the flux with $M_X$ can
be found to be
\begin{eqnarray}
\label{smoothpropto}
\mathcal{F} & \propto & M_X^{1/2} \langle\sigma_A\rangle M_X^{-2} \ \
(\textrm{annihilation}) \nonumber \\
& \propto & M_X^{1/2} \Gamma_X M_X^{-1} \ \ (\textrm{decay}) .
\end{eqnarray}
The factor of $M_X^{1/2}$ is from the fragmentation function, and the factors
of $M_X^{-2}$ or $M_X^{-1}$ arise from $n_X^2$ or $n_X$,
respectively. Therefore, for a given $\mathcal{F}$, the necessary cross section
scales as $M_X^{3/2}$ in the annihilation scenario and the necessary decay
width scales as $M_X^{1/2}$ in the decay scenario.

Calculating the resulting UHE cosmic ray flux in the annihilation model, and
comparing it to the similar calculation in the decay scenario, we obtain the
results shown in Fig.\ \ref{spectrum}.

%%%%%%%%%%%%%%%%%%%%%%%%%%%%%%%%%%%%%%%%%%%%%%%%%%%%%%%
%%%%%%%%%%%%%%%%%%%%%%%%%%%%%%%%%%%%%%%%%%%%%%%%%%%%%%%

The shape of the spectrum is determined by the mass of the \WIMPZILLA\ and the
overall normalization can be scaled by adjusting $\langle\sigma_Av\rangle$ or
$\Gamma_X$. Clearly, in order to produce UHE cosmic rays in excess of
$10^{20}\,$eV, $M_X$ cannot be much smaller than $10^{12}\,$GeV.  In order
to provide enough events to explain the observed UHE cosmic rays,
$\langle\sigma_Av\rangle$ has to be in the range $10^{-25}\,\textrm{cm}^2$ to
$10^{-27}\,\textrm{cm}^2$.

This is well in excess of the unitarity
bound to the $l$-wave reaction cross section \cite{SW,GK,hui}:
\begin{eqnarray*}
\sigma_l v & \le & 
\frac{4\pi}{M^2v}(2l+1)
\simeq \frac{1.5\times 10^{-47}}{M_{12}^2} \nonumber \\
& &  \times (2l+1) \frac{100\,\textrm{km s}^{-1}}{v}\ \textrm{cm}^2 .
\end{eqnarray*}
The unitarity bound essentially states that the annihilation cross section must
be smaller than $M_X^{-2}$.  However, as emphasized by Hui \cite{hui}, there
are several ways to evade the bound.  The annihilation cross section may be
larger if there are fundamental length scales in the problem larger than
$M_X^{-1}$.  One example of a fundamental length scale might be the physical
size of the particle.  In this case, the scale for the annihilation cross
section could be related to the size of the particle.  Another possibility for
the scale of the annihilation cross section might be the range of the
interaction.  If the interaction has a range $\lambda$, then one might imagine
that the scale for the cross section is $\lambda^2$.

A related issue is the typical energy of the annihilation products.  In this
paper we assume that annihilation produces two jets, each with energy
approximately $M_X$.  It is easy to imagine that with the finite-size effects
discussed above, there is the possibility that annihilation will produce many
soft particles, rather than essentially two particles each of energy $M_X$.

An example that suits our needs is the annihilation of a monopole-antimonopole
pair.  Assume the monopole mass is $M_M$.  In the early universe, the
annihilation cross section is roughly $M_M^{-2}$ \cite{Preskill}.  It is this
cross section that would determine the present monopole abundance.  However,
monopole-antimonopole annihilation in the present universe is another matter
\cite{Hill}.  In the late universe, monopoles and antimonopoles can capture 
with a large cross section and form monopolonium, a monopole--antimonopole 
Rydberg state.  It may take as long as the age of the universe for the state to
lose energy by radiation, becoming smaller and more tightly bound.  Eventually
the state will have a size of the inner core of the monopole (for a GUT
monopole, about $\alpha /M_M$).  Then the monopole-antimonopole annihilates in
a final burst of radiation.

The annihilation spectrum of monopolonium contains many particles.  There are
many low-energy photons radiated during the long decay period, followed by a
rapid burst of high-energy jets ($E\sim \alpha M_M$) in the final death throes
of annihilation.

The monopolonium example also illustrates that the early-universe annihilation
cross section may be many orders of magnitude different than the present
annihilation cross section.  
However, for magnetic monopoles the
Parker bound on their abundance
imposes a further constraint \cite{parker}. This arises from the
requirement that magnetic monopoles  
must not cause too much depletion of the galactic
magnetic field and
seems to rule out a noticeable UHE cosmic ray flux from their
annihilation in the halo \cite{olum}.

The monopolonium example is an existence proof that the annihilation cross
section may be many orders of magnitude larger than $M_X^{-2}$, while
annihilation products can have energy of order $M_X$.  While it is an existence
proof, it also illustrates that the combined conditions of $\sigma_A \gg
M_X^{-2}$ and $\langle E \rangle \sim M_X$ probably requires some unusual
interactions of the dark matter. Of course we know essentially nothing about
the interactions of dark matter, in particular its self-interactions, which may
not be communicated to the visible sector.  
 
We regard the requisite size of the
annihilation cross section to be the most unattractive feature of our proposal.

%In the next section we will show that the signal from
%isothermal clumps of dark matter may explain the observed flux
%without violating unitarity bounds. 

%%%%%%%%%%%%%%%%%%%%%%%%%%%%%%%%%%%%%%%%%%%%%%%%%%%%%%%
%%%%%%%%%%%%%%%%%%%%%%%%%%%%%%%%%%%%%%%%%%%%%%%%%%%%%%%
\section{Annihilation in the clumped component}
\label{clumped}
%%%%%%%%%%%%%%%%%%%%%%%%%%%%%%%%%%%%%%%%%%%%%%%%%%%%%%%
%%%%%%%%%%%%%%%%%%%%%%%%%%%%%%%%%%%%%%%%%%%%%%%%%%%%%%%

So far we have assumed that the galactic dark matter is smoothly distributed.
In this section we will consider the contribution from inhomogeneities in the
galactic distribution.  

We have modeled the smooth component of the dark matter by a NFW profile, Eq.\
(\ref{nfwp}).  In addition to this smooth component, the dark matter may have a
clumped component, as suggested by $N$-body simulations. Using the results of
these simulations, the number of subclumps of mass $M_{cl}$ per unit volume at
distance $d$ from the galactic center can be written in the form:
\begin{eqnarray*}
n_{cl}(d,M_{cl}) & = & n_{cl}^0 \left(\frac{M_{cl}}{M_H}\right)^{-\alpha}
\nonumber \\  & & \times \left[1+
\left(\frac{d}{R^{cl}_c}\right)^2\right]^{-3/2},
\end{eqnarray*}
where $R^{cl}_c$ is the core radius of the subclump distribution in the galaxy,
typically of order 10 to 20 kpc. The mass of the halo of the Galaxy is
$M_H$. The normalization constant $n_{cl}^0$ can be calculated by requiring
that the mass in the clumped component is a fraction $\xi$ of the halo mass
$M_H$. From simulations, $\xi\sim 10\%$.  The power index, $\alpha$, may also
be found from simulations, with the result $\alpha \sim 1.9$. It is then easy
to find
\[
n_{cl}^0 \approx \frac{(2-\alpha)\xi}{4\pi (R^{cl}_c)^3\ln(R_H/R^{cl}_c)M_H
\eta^{2-\alpha}},
\]
where $\eta$ is the ratio of the mass of the largest subclump to the halo mass,
expected to be in the range 0.01 to 0.1, and $R_H$ is the radius of the halo.

Up to this point, all the discussion is independent of the density profile of
the individual subclumps. 

Let us now assume that the density profile of an individual subclump is an
isothermal sphere, with radial profile
\begin{equation}
n_X(r)=n_H(d)\left(\frac{r}{r_0}\right)^{-2},
\label{eq:isotherm}
\end{equation}
where $r_0$ is the radius of the individual subclump, defined as the radius at
which the subclump density equals the density of the background halo at the
distance $d$ from the galactic center. 

If the subclump mass is $M_{cl}$, we can write
\[
M_{cl} = 4\pi  M_X n_H(d)\int_0^{r_0} dr \,  r^2 (r/r_0)^{-2},
\]
from which we obtain
\begin{equation}
r_0 = \left[ \frac{M_{cl}}{4\pi M_X n_H(d)}\right]^{1/3} .
\label{eq:clumpradius}
\end{equation}

The rate of annihilation in an individual subclump at distance $d$ from the
galactic center is
\begin{eqnarray}
\mathcal{R}(M_{cl},d) & = &  4\pi n^2_H(d)\langle\sigma_Av\rangle r_0^4
\nonumber \\
& & \times \left(\frac{1}{3R_{min}}+
\int_{R_{min}}^{r_0}\frac{dr}{r^2}\right)\\
 & \approx & 
\frac{16\pi}{3}n^2_H(d)\langle\sigma_Av\rangle \frac{r_0^4}{R_{min}},
\nonumber 
\label{eq:lumi}
\end{eqnarray}
where $R_{min}$ is the minimum radius of a subclump (the inner radius where the
profile is flattened by efficient annihilations).

One way to estimate $R_{min}$ would assume equality of the annihilation
time scale
\begin{eqnarray*}
\tau_A & = &\left[n(r) \langle\sigma_Av\rangle\right]^{-1} \nonumber \\
& = & \frac{1}{n_H(d) \langle\sigma_Av\rangle} \left(\frac{r}{r_0}\right)^2
\end{eqnarray*}
with the radial infall time
\begin{eqnarray*}
\tau_{ff} & = & r_0/v(r_0) \nonumber \\
& = & \left[4\pi G M_X n_H(d) \right]^{-1/2},
\end{eqnarray*}
see e.g. \cite{Ber1}, where such an estimate was applied to the central core
of the galaxy.

In the present setting this would yield a ratio
\begin{equation}
\left.\frac{R_{min}}{r_0}\right|_{\tau_{ff}=\tau_A} = 
\langle\sigma_Av\rangle^{1/2} 
\left[\frac{n_H(d)}{4\pi G M_X} \right]^{1/4}
\label{eq:rmin}
\end{equation}
and an annihilation rate
in a subclump of mass $M_{cl}$ at distance $d$ from the
galactic center:
\begin{eqnarray*}
\mathcal{R}(M_{cl},d) & = & \frac{4}{3}
 M_{cl} \langle\sigma_Av\rangle^{1/2} 
(4\pi G)^{1/4} \nonumber \\
& & \times \left(\frac{n_H(d)}{M_X}\right)^{3/4} .
\end{eqnarray*}

Now we can calculate the contribution of the clumped component in a similar
manner as Eq.\ (\ref{smoothannihilation}):
\begin{eqnarray}
\label{clumpedisoannihilation}
\mathcal{F} & = &
 \frac{d\mathcal{N}(E,M_X)}{dE} \int d^3\vek{d} \ 
\frac{1}{2\pi \left|\vek{d}-\vek{d}_\odot\right|^2} 
\nonumber \\
& & \times \int_{M_{min}}^{M_{max}} dM_{cl} \
 n_{cl}(d,M_{cl})\mathcal{R}(d,M_{cl}).
\end{eqnarray}
As before we will use an NFW profile for $n_H(d)$ with $d_s=24\,$kpc.

Using
$R_c^{cl}=15\,$kpc, $\eta=0.1$, and $\xi = 0.1$, we find
\begin{eqnarray*}
\frac{\mathcal{F}_\textrm{clumped}}{\mathcal{F}_\textrm{smooth}} & \simeq & 
2\times10^6 \left(\frac{M_X}{10^{12}\textrm{GeV}}\right)^{1/2}
\\ & & \times \left(
\frac{3\times10^{-26}\,\textrm{cm}^2}{\langle\sigma_Av\rangle}\right)^{1/2}.
\end{eqnarray*}

To understand the scaling of the events from subclumps, we can make the simple
approximation that one is at the galactic center ($|\vek{d}_\odot|=0$).  In the
limit that the contributions come from $d\ll R^{cl}_c$ and $d\ll d_s$, we find
\[
\mathcal{F} \propto \langle\sigma_Av\rangle^{1/2} M_X^{-1} .
\]

Now, we take the density profile of the subclumps according to a NFW profile,
\begin{equation}
n_X(r)=n_H(d)\,\frac{r_0\left[r_0+r_s\right]^2}{r\left[r+r_s\right]^2},
\end{equation}
where $r_s$ is the fiducial radius of the subclump and again $r_0$ is the
radius of the subclump where $n_X(r)=n_H(d)$. We
assume here that the ratio $r_0/r_s\gg 1$ is constant for all subclumps.
 The mass of the subclump can be written in terms of $r_0$
and $r_s$ as
\begin{eqnarray*}
M_{cl} & = & 4\pi M_X \int_0^{r_0} dr \,  r^2 n_X(r) \nonumber \\ 
& \simeq & 4\pi M_X n_H(d)r_0^3\ln\left(\frac{r_0}{r_s}\right).
\end{eqnarray*}
which provides the core radius $r_0$ for the subclump once we know $r_s$.

The rate of annihilations for one subclump is given by 
\begin{eqnarray*}
\mathcal{R}(M_{cl},d) & = & 4\pi \langle\sigma_Av\rangle  
\int_{R_{min}}^{r_0} dr\, r^2 n_X^2(r)\\
& & +(4\pi/3)R_{min}^3\langle\sigma_Av\rangle n^2_H(R_{min}),\\
\label{eq:lumiNFW}
\end{eqnarray*}
which becomes
\[
\mathcal{R}(M_{cl},d)\simeq \frac{4\pi}{3} \langle\sigma_Av\rangle 
\frac{r_0^6}{r_s^3}n_H^2(d) .
\]

\begin{figure}[htb]
\begin{center}
\resizebox{0.5\textwidth}{!}{ \includegraphics{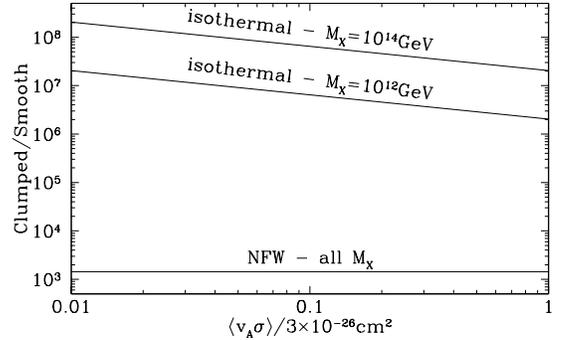} }
\end{center}
\caption{The ratio of events from the subclump component to the smooth
component assuming either an isothermal or a NFW profile for the subclumps.
This assumes the ratio (\ref{eq:rmin})}
\label{clumpedfig}     
\end{figure}

To calculate $\mathcal{F}$ for NFW subclumps, we follow a procedure similar to 
Eq.\ \ref{clumpedisoannihilation}, but with $\mathcal{R}(M_{cl},d)$ given
above.  Again expressing the flux in terms of the flux from the smooth
component, we find 
\[
\frac{\mathcal{F}_\textrm{clumped}}{\mathcal{F}_\textrm{smooth}} \simeq
2\times10^3 ,
\]
independent of $M_X$ or $\langle\sigma_Av\rangle$.  Again, this result is
represented in Fig.\ \ref{clumpedfig}.

%%%%%%%%%%%%%%%%%%%%%%%%%%%%%%%%%%%%%%%%%%%%%%%%%%%%%%%
%%%%%%%%%%%%%%%%%%%%%%%%%%%%%%%%%%%%%%%%%%%%%%%%%%%%%%%

The ratio of the fluxes from the subclump component and the smooth component
are given in Fig.\ \ref{clumpedfig}.  Comparing Fig.\ \ref{spectrum} and Fig.\
\ref{clumpedfig}, we see that for $M_X=10^{12}\,$GeV, to normalize the flux to
the observed events would require a cross section $10^3$ times smaller than the
isotropic case if the \WIMPZILLA\ subclumps have an NFW profile.

For NFW subclumps the calculation of the flux has a natural cutoff at small
$r$. This is not the case for isothermal subclumps, and the resulting flux
could be very different with a different choice for $R_{min}$.  For example,
one might estimate $R_{min}$ by requiring detailed hydrodynamic balance between
the infalling and the annihilating matter in the central core of the dark
matter clumps. In general we can write for the change in density of a radially
symmetric dark matter clump
\begin{eqnarray}\label{eq:evolve}
\frac{\partial}{\partial t} n_X(r,t) & = & \frac{\partial}{\partial r}
(n_X(r,t) v(r,t))  \\
& & + \frac{2}{r} n_X(r,t)v(r,t) \nonumber \\
& & - n_X^2(r,t) \langle\sigma_A v\rangle,
\nonumber
\end{eqnarray}
where $v(r,t)$ is the velocity of radial infall.

Detailed balance at $r=R_{min}$ amounts to the requirement
\begin{equation}
v(R_{min})=\frac{1}{3}R_{min}n_X(R_{min})\langle\sigma_Av\rangle.
\label{eq:balance}
\end{equation}
Such a requirement was applied in \cite{Ber2} to estimate the core radii
of neutralino stars.

For a particle of energy $E=0$ we find from energy conservation
or from the quasi-stationary
Euler equation in \cite{Ber2}
\[
v^2(r) = -2U(r),
\]
where the potential $U(r)$ in the clump can be calculated 
as a solution of the  Poisson equation:
\begin{eqnarray}\label{eq:potential}
U(r) & = & \frac{2\pi GM_X}{r}\\
 & & \times \int_0^{r_0}dr'\,n_X(r') r'(|r-r'|-r-r').
\nonumber
\end{eqnarray}
This yields
\[
U(R_{min})\approx -4\pi GM_X n_H(d)r_0^2\ln\!\left(\frac{r_0}{R_{min}}\right)
\]
and amounts to a ratio
\[
\left.\frac{R_{min}}{r_0}\right|_{balance}\approx 0.1\langle\sigma_Av\rangle
\left[\frac{n_H(d)}{72\pi GM_X}\right]^{1/2},
%\label{eq:rmin2}
\]
much smaller than what found in Eq.\ (\ref{eq:rmin}).  With this estimate the
rate of annihilation in an individual subclump at distance $d$ from the
galactic center is
\begin{equation}
\mathcal{R}(d,M_{cl})\approx 80 M_{cl}
\left[\frac{2\pi G n_H(d)}{M_X}\right]^{1/2},
\label{eq:lumi2}
\end{equation}
and the flux is
\begin{eqnarray*}
\mathcal{F}_{\textrm{\small clumped}}
 & \approx & \frac{d\mathcal{N}(E,M_X)}{dE}\times 4.2\,\mbox{m}^{-2}\,
\mbox{s}^{-1}\\
 & & \times\left(\frac{\xi}{0.1}\right).
\end{eqnarray*}
This is by about a factor $10^{15}$ {\em too large}, and can
be reconciled with the observed UHE cosmic ray spectrum only if the
\WIMPZILLA\ contribution to $\Omega_{dark}$  amounts only
to a small admixture $\sim 10^{-15}$ of annihilating
\WIMPZILLAS\ in sub-galactic dark matter clumps.

Note that this result is independent of the annihilation cross section
$\langle\sigma_A v\rangle$, and therefore also complies with annihilation cross
sections well below the $s$-wave unitarity limit. With the estimate
(\ref{eq:balance}) a small annihilation cross section implies a very small
densely packed core region of the dark matter clumps. The high density
increases the luminosity of the clumps and compensates for the small factor
$\langle\sigma_A v\rangle$ in the local annihilation rate.

%%%%%%%%%%%%%%%%%%%%%%%%%%%%%%%%%%%%%%%%%%%%%%%%%%%%%%%
%%%%%%%%%%%%%%%%%%%%%%%%%%%%%%%%%%%%%%%%%%%%%%%%%%%%%%%
\section{Predicted signals}
\label{anisotropy}
%%%%%%%%%%%%%%%%%%%%%%%%%%%%%%%%%%%%%%%%%%%%%%%%%%%%%%%
%%%%%%%%%%%%%%%%%%%%%%%%%%%%%%%%%%%%%%%%%%%%%%%%%%%%%%%

The annihilation rate is very sensitive to the local density.  This implies
that if the UHE cosmic rays originate from dark matter annihilation and dark
matter in our galaxy is clumped, the arrival direction of UHE cosmic rays
should reflect the dark matter distribution.

\begin{figure}[htb]
\begin{center}
\resizebox{0.5\textwidth}{!}{ \includegraphics{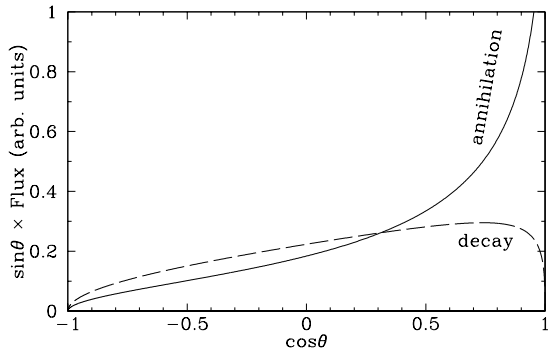} }
\end{center}
\caption{The angular dependence of events originating from annihilation or
decay of the smooth component of dark matter that follows a NFW profile.  In
this figure, $\cos\theta=1$ corresponds to the direction of the galactic
center.}
\label{angle}     
\end{figure}

The first possibility we considered is that the dark matter distribution is
smooth and follows a NFW profile.  In this case, the galactic center should be
quite prominent.  In Fig.\ \ref{angle} we illustrate the expected angular
dependence of the arrival direction.  The galactic center is prominent for the
decay scenario where $\mathcal{F}\propto n_X$ \cite{bbv}, and 
even more prominent for the
annihilation case where $\mathcal{F}\propto n_X^2$.

%%%%%%%%%%%%%%%%%%%%%%%%%%%%%%%%%%%%%%%%%%%%%%%%%%%%%%%
%%%%%%%%%%%%%%%%%%%%%%%%%%%%%%%%%%%%%%%%%%%%%%%%%%%%%%%

Now if we assume that there are isothermal
dark matter subclumps within the galaxy, then
the density in the subclumps would be larger than the ambient background
dark-matter density, and events originating from subclumps will dominate the
observed signal.

The dominance of events from subclumps has two effects.  The first effect is
that a smaller annihilation cross section is required to account for the
observed UHE flux.  The second effect is that there will be a very large
probability of detecting a nearby subclump.

%%%%%%%%%%%%%%%%%%%%%%%%%%%%%%%%%%%%%%%%%%%%%%%%%%%%%%%
\begin{figure}[htb]
\begin{center}
\resizebox{0.5\textwidth}{!}{ \includegraphics{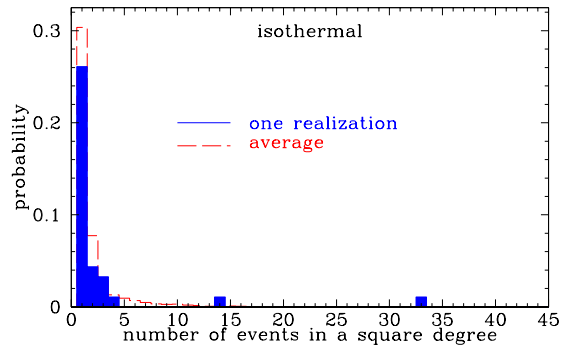} }
%\resizebox{0.5\textwidth}{!}{ \includegraphics{clumpednfw.eps} }
\end{center}
\caption{The probability of finding a certain number of events in a one square
degree area of the sky}  
%The upper figure is for isothermal subclumps and the
%lower figure for NFW subclumps}
\label{subclumpdistribution}     
\end{figure}
%%%%%%%%%%%%%%%%%%%%%%%%%%%%%%%%%%%%%%%%%%%%%%%%%%%%%%%

In Fig.\ \ref{subclumpdistribution} we present a histogram showing the number
of occurrences of single and multiple events in arrival directions within a
square degree.  We generated many realizations of the expected subclump
distribution, and calculated the flux from the subclumps that would result in
about 100 detected events. We see that the {\it average} probabilities are
quite reasonable, with most events in square degree areas with only one event,
and a few pairs and triplets of events.  However if we examine individual
realizations containing about 100 events, we see that there is a large
probability of a large number of events from single nearby subclumps.  For
instance, in the single realizations shown in Fig.\
\ref{subclumpdistribution}, the isothermal subclumps result in a square degree
bin with 14 events and a square degree bin with 33 events.  
%In the typical NFW
%distribution, there is one bin with 16 events and another bin with 43 events.
%Therefore, 
If more than 100 events are observed with full sky coverage, the
signature of subclumps will be unmistakable.

%%%%%%%%%%%%%%%%%%%%%%%%%%%%%%%%%%%%%%%%%%%%%%%%%%%%%%%
%%%%%%%%%%%%%%%%%%%%%%%%%%%%%%%%%%%%%%%%%%%%%%%%%%%%%%%
\section{Conclusion}
\label{conclusion}
%%%%%%%%%%%%%%%%%%%%%%%%%%%%%%%%%%%%%%%%%%%%%%%%%%%%%%%
%%%%%%%%%%%%%%%%%%%%%%%%%%%%%%%%%%%%%%%%%%%%%%%%%%%%%%%

The origin of UHE cosmic rays from annihilation (or decay) of \WIMPZILLAS\ has
the attractive feature of the simplicity of generating ultrahigh energies:
simple conversion of rest mass energy.  

UHE cosmic rays from decay of a smooth relic superheavy dark matter population in the
halo requires a lifetime of about $10^{20}\,$years or a correspondingly
reduced contribution of the superheavy relics to $\Omega_{dark}$.

In this paper we have examined the annihilation scenario:
We found that unitarity limits very likely exclude an appreciable contribution
to UHE cosmic rays from standard particle annihilation in the smooth halo
component, but monopolonium decay is
a possible counterexample to this conclusion.

We also found that annihilation 
of \WIMPZILLA\ particles in
sub-galactic clumps does not necessarily violate low-$l$ unitarity limits.  
Like the decay of \WIMPZILLAS\ of lifetime $\tau\sim 10^{10}\,$years,
this would also imply
$\Omega_\mathrm{\WIMPZILLA}\ll\Omega_{dark}$.

The annihilation proposal
results in several striking predictions.  While not discussed in this paper,
the annihilation scenario suggests that the bulk of the UHE events are 
photons. This prediction, common to the decay scenario, seems to be 
already in conflict with a recent analysis of the composition of the 
old Haverah Park data, which indicate that no more than $55\%$ of events 
above $4\times 10^{19}$ eV are consistent with being photons \cite{zas2000}. 
However, this limit 
is weaker for events with energy above $10^{20}$ eV. 
Upcoming
data with the large statistics of events required to determine the 
composition beyond any doubt will soon be available and will definitely
test the viability of all the top down models of UHECRs.

The true characteristic signature of the annihilation
scenario is the expected anisotropy in arrival direction.  If the dark matter
is smoothly distributed in the galaxy, the galactic center should be prominent.
If the dark matter is clumped on sub-galactic scales, then the subclumps should
be visible.

Thus, the annihilation scenario can be falsified by complete sky coverage.  The
Pierre Auger Observatory and future space based experiments, such 
as EUSO \cite{scarsi} and OWL \cite{nasa} will be able to see the 
galactic center, and, by covering the southern sky, should be able 
to pick out subclumps if they are present.\\[1ex]

%%%%%%%%%%%%%%%%%%%%%%%%%%%%%%%%%%%%%%%%%%%%%%%%%%%%%%%
%%%%%%%%%%%%%%%%%%%%%%%%%%%%%%%%%%%%%%%%%%%%%%%%%%%%%%%
{\bf Acknowledgement:}\\ R.D.\ thanks Julio Navarro for a conversation on the
NFW halo profile.  P.B.\ and E.W.K.\ would like to thank Keith Ellis for
discussion about QCD hadronization.  The work of P.B.\ and E.W.K.\ was
supported in part by NASA (NAG5-7092). The work of R.D.\ was supported
in part by NSERC Canada.
%%%%%%%%%%%%%%%%%%%%%%%%%%%%%%%%%%%%%%%%%%%%%%%%%%%%%%%
%%%%%%%%%%%%%%%%%%%%%%%%%%%%%%%%%%%%%%%%%%%%%%%%%%%%%%%

%%%%%%%%%%%%%%%%%%%%%%%%%%%%%%%%%%%%%%%%%%%%%%%%%%%%%%%
%%%%%%%%%%%%%%%%%%%%%%%%%%%%%%%%%%%%%%%%%%%%%%%%%%%%%%%

%%%%%%%%%%%%%%%%%%%%%%%%%%%%%%%%%%%%%%%%%%%%%%%%%%%%%%%
%%%%%%%%%%%%%%%%%%%%%%%%%%%%%%%%%%%%%%%%%%%%%%%%%%%%%%%
\end{document}